\documentclass[aps,pra,twocolumn,superscriptaddress,amsmath,amssymb]{revtex4-2}
\usepackage{graphicx}
\usepackage{subfigure} 
\usepackage[colorlinks=true]{hyperref}
\usepackage{cancel}

\usepackage{ulem}
\newcommand{\wzy}[1]{\textcolor{black}{#1}}
\newcommand{\why}[1]{\textcolor{black}{#1}}

\begin{document}

\title{Enhancing the fidelity of stimulated Raman transitions with simple phase shifts}

\author{Hua-Yang Wang}
\affiliation {Key Laboratory of Atomic and Subatomic Structure and Quantum Control (Ministry of Education), Guangdong Basic Research Center of Excellence for Structure and Fundamental Interactions of Matter, School of Physics, South China Normal University, Guangzhou 510006, China}

\author{Xu-Yang Chen}
\affiliation {Key Laboratory of Atomic and Subatomic Structure and Quantum Control (Ministry of Education), Guangdong Basic Research Center of Excellence for Structure and Fundamental Interactions of Matter, School of Physics, South China Normal University, Guangzhou 510006, China}

\author{Zhen-Yu Wang}
\email{zhenyu.wang@m.scnu.edu.cn}
\affiliation {Key Laboratory of Atomic and Subatomic Structure and Quantum Control (Ministry of Education), Guangdong Basic Research Center of Excellence for Structure and Fundamental Interactions of Matter, School of Physics, South China Normal University, Guangzhou 510006, China} 
\affiliation {Guangdong Provincial Key Laboratory of Quantum Engineering and Quantum Materials, Guangdong-Hong Kong Joint Laboratory of Quantum Matter, South China Normal University, Guangzhou 510006, China}

\begin{abstract}
We demonstrate that in stimulated Raman transitions, introducing one or two simple phase shifts to the control fields significantly enhances the fidelity of state manipulation while simultaneously reducing leakage to the intermediate excited state. Our approach achieves high-fidelity quantum gate operations between the two target states under arbitrary detuning conditions. Notably, the average population in the intermediate excited state is approximately halved, without extending the overall evolution time. Additionally, our method exhibits greater robustness to static amplitude and detuning errors compared to \wzy{conventional} adiabatic elimination techniques, and maintains higher fidelity even in the presence of dissipation.
\end{abstract}

\maketitle

\section{Introduction}
 	High-fidelity quantum state manipulation is crucial for applications such as quantum simulation, quantum sensing, and quantum computing~\cite{Saffman2010RMP, Georgescu2014, Degen2017RMPsensing}. In many cases, however, direct coupling between quantum states is not possible due to forbidden transitions. Indirect coupling can be achieved through intermediate states~\cite{Gaubatz1990, Bergmann1998, Vasilev2009, Vitanov2017, Moler1992, Shkolnikov2020}. For example, transitions between two ground states without a direct coupling pathway can be realized by using two polarized lasers that connect these ground states via a common intermediate excited state in a $\Lambda$-type configuration~\cite{Golter2014}. To enhance the fidelity of state manipulation, it is important to minimize the population of the excited state, thus mitigating the negative impacts of dissipation and spontaneous decay.
	
Stimulated Raman adiabatic passage (STIRAP) has been widely studied and implemented experimentally~\cite{Gaubatz1990,Bergmann1998,Vitanov2017,Vitanov1997, Kumar2016,Golter2014,Goto2006,Klein2007,Du2014,Fedoseev2021,Bohm2021}. 
STIRAP, which is an adiabatic scheme, can be used to transfer the ground states without direct transition by slowly changing the relative strengths of the two partially overlapping control fields which couple the two target states to an intermediate excited state~\cite{Vitanov1999fractionalSTIRAP,Vitanov2017}. With the transfer state being a dark state,
this adiabatic scheme suppresses the population of the intermediate excited state during the adiabatic evolution.
As a consequence, the scheme is not affected by the dissipation and spontaneous decay of the excited state. 
As an adiabatic scheme, STIRAP offers some robustness to control errors such as the intensity and shape of the control field~\cite{Vitanov2017} and its performance can be improved by adjusting the waveform of control field~\cite{Vasilev2009,Verdeny2014}. 
However, to satisfy the adiabatic condition, STIRAP requires a long control time, which can suffer from the problem of decoherence.

The adiabatic elimination (AE)~\cite{Moler1992,Fewell2005, Brion2007,Paulisch2014} of the stimulated Raman transitions (SRT) is another conventional three-level system control scheme. In AE, the two control fields, which couple the two states for manipulation in a two-photon resonance manner, induce transitions between the two states via a second order effect. The effects of dissipation and spontaneous decay in the intermediate excited state can be suppress by using a large single-photon detuning, because in AE the population of the intermediate state is inversely proportional to the square of this detuning.
However, a large detuning also implies a weak transition between the two manipulated states because 
their effective coupling is inversely proportional to the detuning due to the second-order effect. This makes the control on the ground state manifold slow, and environmental noise can have enough time to spoil the ground state manifold.
 
In order to accelerate the state evolution, shortcuts to adiabaticity (STA)~\cite{Demirplak2003,Demirplak2005,Demirplak2008,Berry2009,Chen2010STA,
Ibanez2012STA,Wang2016necessary,Zhou2017Superadiabatic,
Odelin2019,Xu2019QSLGeodesics,Zheng2022Geodesics,Liu2022shortcuts,
Gong2023Geodesics,Gong2024Superadiabatic,chen2024fast} has been developed. The idea of STA is to achieve the same state transfer in quantum adiabatic evolution, but with a shorter evolution time. For example, the counterdiabatic driving \cite{Demirplak2003,Demirplak2005,Demirplak2008,Berry2009,Chen2010STA,Campo2013} eliminates the non-adiabatic transitions between the instantaneous eigenstates of the original Hamiltonian by adding an auxiliary control Hamiltonian. However, the auxiliary control Hamiltonian may not be feasible in experiments. For example, STA in general requires unavailable direct transition between the ground states of quantum systems of $\Lambda$-type configuration~\cite{Chen2010STA}. Alternatively, one can apply the STA technique to the effective two-level system obtained after the use of AE~\cite{Li2016,Du2016}, which, however, also encounters the problems of slow control speed and decoherence. In addition, 
adding the auxiliary control Hamiltonian inevitably modifies the total Hamiltonian, which implies that the evolution of quantum states no longer follows the instantaneous eigenstates of the total Hamiltonian, and therefore for STA the robustness against control errors is not guaranteed~\cite{Torosov2021}. 
	
	In this work, we present a simple scheme that enhances AE by incorporating one or two phase shifts into the control fields. Adding a simple phase shift at the beginning of the AE control fields enables a substantial reduction in intermediate state occupation, while maintaining the overall control time. By introducing a second phase shift and reversing the detuning at the end of the control process, we can eliminate leakage to the intermediate state entirely in the absence of decay. \wzy{The underlying physical principle of our protocol is the use of coherent cancellation during state evolution to reduce the population in the intermediate state, somewhat similar to coherent averaging techniques\why{~\cite{Suter2016RMP,Hayes2012}}.} We demonstrate that our approach outperforms both the AE and STIRAP methods, offering greater robustness against static amplitude and detuning errors, as well as improved resilience to decoherence. \wzy{Our scheme is highly feasible for experimental implementation, as it requires only simple adjustments to the phase and detuning. Finally, we show a simple example of our method to reduce leakage in two-qubit gates.}

\section{Leakage to the intermediate state}

\begin{figure}[]
		\centering
		\includegraphics[width=\columnwidth]{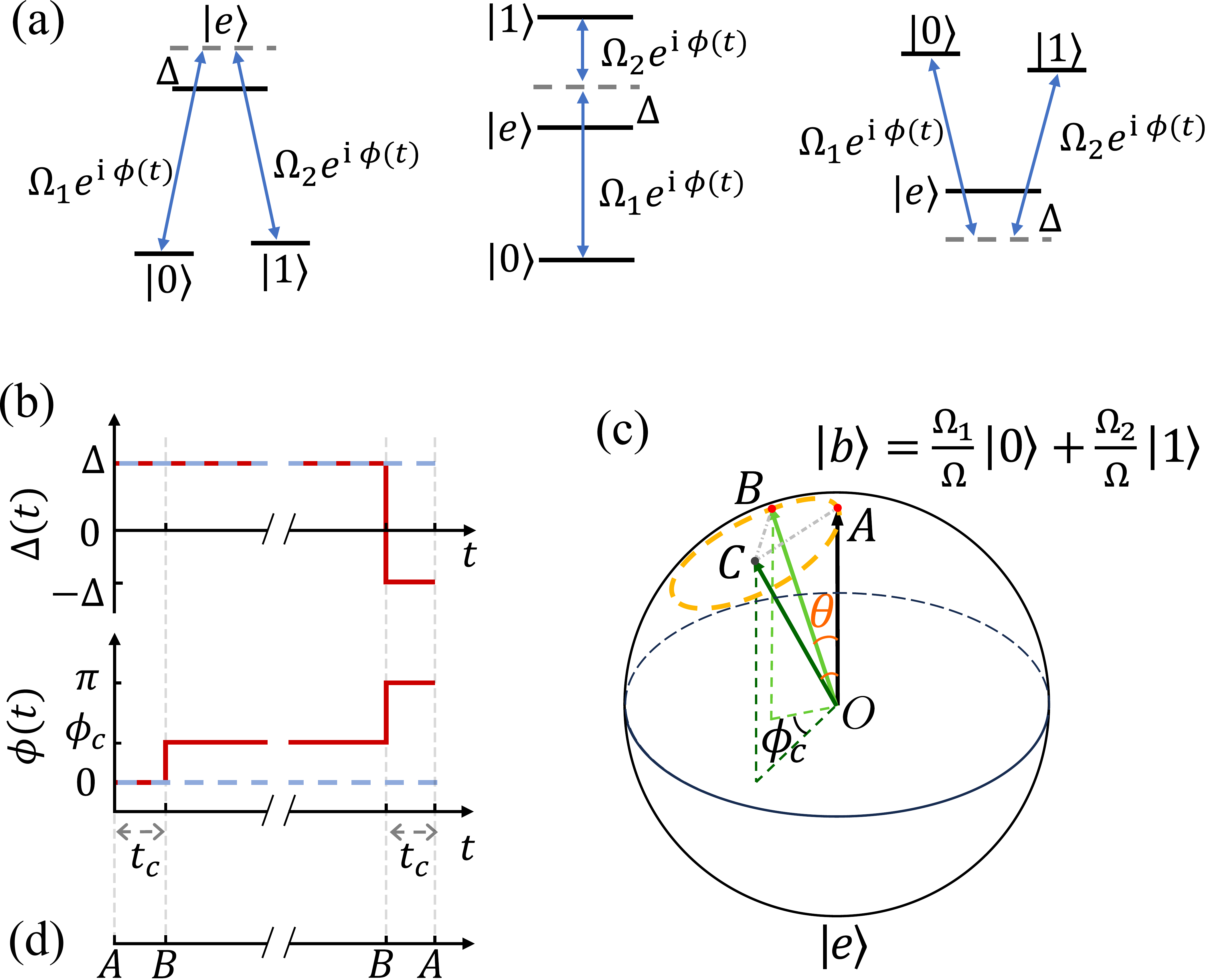}
		\caption{(a) Indirect transitions between states $|0\rangle$ and $|1\rangle$ via coupling to an intermediate state $|e\rangle$ in a variety of three-level configurations. (b) Time evolution of the detuning $\Delta(t)$ and the common control phase $\phi(t)$ for AE (blue dashed lines) and our proposed \wzy{scheme} (red lines). (c) Illustration of the \wzy{state evolution starting from the bright state $|b\rangle$, which is indicated by Point $A$ on the Bloch sphere spanned by $\{|b\rangle,|e\rangle\}$.} The yellow dashed line shows the \wzy{state evolution trajectory under a constant AE Hamiltonian with $\phi(t)=0$ corresponding to Point $C$. 
At the moment $t_c$, the state evolves to Point $B$, which is equivalent to Point $C$ up to a rotation around the $OA$ axis (hence Points $A$, $B$, and $C$ form an equilateral triangle).
Using an immediate change of the phase to $\phi(t)=\phi_c$ in our scheme, the state at Point $B$ becomes stationary because it is an eigenstate of the Hamiltonian with $\phi(t)=\phi_c$.  For conventional AE the state will pass Point $B$ and has a larger leakage to $|e\rangle$.
(d) The trajectory points of our scheme. }
}		\label{fig:fig1}
\end{figure}

	To better understand the limitations of conventional methods and introduce our approach, we consider a three-level system in which the two states $|0\rangle$ and $|1\rangle$ do not have a direct transition — see Fig.~\ref{fig:fig1}(a).
\wzy{In Sec.~\ref{sec:qubit}, we provide an example of the three-level transition model.}
Two control fields are applied, each coupling one of \wzy{$|0\rangle$ and $|1\rangle$} to \wzy{the intermediate state} $|e\rangle$ under the two-photon resonance condition. The Hamiltonian of the system in the rotating frame reads  (with $\hbar=1$)
\begin{align}
H(t)=-\Delta\wzy{(t)}|e\rangle\langle e|+ \left[\frac{\Omega}{2}e^{i\phi\wzy{(t)}}|b\rangle\langle e|+\text{H.c.}\right],
\label{eq:systemHamiltonian}
\end{align}
where $\Delta\wzy{(t)}$ is the single-photon detuning and $\phi\wzy{(t)}$ is a common phase of the control field. 
\wzy{
In the conventional AE method discussed in the paper, $\Delta(t)=\Delta$ and $\phi(t)=\phi_0$ (e.g., $\phi_0=0$) are constant parameters, whereas in our method, they are time-dependent, see Fig.~\ref{fig:fig1}(b).}  Here
we define the bright state 
\begin{align}|b\rangle=\frac{1}{\Omega}(\Omega_1|0\rangle+\Omega_2|1\rangle),
\end{align}
where $\Omega_1$ ($\Omega_2$) is the (complex) coupling amplitude of the control field for the $|0\rangle$ ($|1\rangle$) and $|e\rangle$ transition. $\Omega=\sqrt{|\Omega_1|^2+|\Omega_2|^2}$ is the normalization constant. 

It is convenient to write the Hamiltonian \why{$H(t)$ in the rotating frame of} 
\begin{align}
U_0(t)=\exp\left[i\int_{0}^{t}\frac{\Delta(t^{\prime})}{2}(|b\rangle\langle b|+|e\rangle\langle e|)dt^{\prime}\right]
\end{align}
as
\begin{align}
\tilde{H}(t)=\frac{\Delta\why{(t)}}{2}(|b\rangle\langle b|-|e\rangle\langle e|)+\left[\frac{\Omega}{2}e^{i\phi\why{(t)}}|b\rangle\langle e|+\text{H.c.}\right], \label{eq:HTilde}
\end{align}
which has the eigenvalues of $\pm\omega/2$ with the corresponding eigenstates $|\varphi_{\pm}(t)\rangle$, and the eigenvalue $0$ for the dark state $|d\rangle$. Here $\omega=\sqrt{\Delta^2+\Omega^2}$ in general is time-dependent but $\omega$ is chosen to be time-independent in the following discussion on AE and our scheme.

Any initial state $|\psi(0)\rangle$ which is a superposition of the ground states $|0\rangle$ and $|1\rangle$ can be written as $|\psi(0)\rangle=\alpha|b\rangle+\beta|d\rangle$,
by the use of the dark sate 
\begin{align}
|d\rangle=\frac{1}{\Omega}(\Omega_2^*|0\rangle-\Omega_1^*|1\rangle).
\end{align}
Since $H\wzy{(t)}$ has no effect on the dark state $|d\rangle$, the state at a later time $t$ is given by
\begin{align}
|\psi(t)\rangle=U(t)|\psi(0)\rangle=\alpha  U_0(t)|\psi_{b,e}(t)\rangle+\beta|d\rangle, \label{eq:psiT}
\end{align}
where  $U(t)=\mathcal{T}e^{-i\int_{0}^{t}H(t^\prime)dt^\prime}$ with $\mathcal{T}$ being the time-ordering operator, and 
\begin{align}
|\psi_{b,e}(t)\rangle =\mathcal{T}e^{-i\int_{0}^{t}\tilde{H}(t^\prime)dt^\prime}|b\rangle, \label{eq:psiBE}
\end{align}
becomes a superposition of the states $|e\rangle$ and $|b\rangle$ due to the evolution driven by the Hamiltonian Eq.~\eqref{eq:HTilde}.

To see the leakage during the method of AE, for simplicity, we consider constant real parameters $\Delta(t)=\Delta$, and $\phi(t)=0$, see Fig.~\ref{fig:fig1}(b). The analysis for $\Omega_1\neq \Omega_2$ is similar. From Eqs.~\eqref{eq:psiT}, \wzy{\eqref{eq:psiBE}}, and \eqref{eq:HTilde}, one can see that the state evolution is within the subspace spanned by $|b\rangle$ and $|e\rangle$. As shown in Fig.~\ref{fig:fig1}(c), $|b\rangle$ rotates at the angular frequency $\omega$ about the rotation axis with the angle $\theta=\arctan(\Omega/\Delta)$ corresponding to the Hamiltonian $\tilde{H}\wzy{(t)}$ on the Bloch sphere spanned by $\{|b\rangle,|e\rangle\}$. The state evolution trajectory is shown by the yellow dashed line. At the regime of large detuning $\Delta\gg\Omega$ (i.e., $\theta\approx 0$) which is required in AE, the average population in the intermediate state $|e\rangle$ is small.
In particular, Eq.~\eqref{eq:psiT} for AE control with $\Delta\gg\Omega$ reads
\begin{align}
|\psi_{\text{AE}}(t)\rangle \approx e^{-i H_{\text{eff}} t}|\psi(0)\rangle - \alpha \sin\left(\frac{\omega t}{2}\right)\frac{i e^{-i \frac{\omega t}{2}} \Omega}{\Delta} |e\rangle, \label{eq:AEPsi}
\end{align}
with the effective Hamiltonian
\begin{align}
H_{\text{eff}} = \frac{\omega-\Delta}{2}|b\rangle\langle b|+0|d\rangle\langle d|, \label{eq:HEff}
\end{align}
for the control on the subspace spanned by $|0\rangle$ and $|1\rangle$.

Equation~\eqref{eq:AEPsi} shows that the population of $|e\rangle$ is 
$P_{e}^{\text{AE}}=\frac{\Omega^2}{\Delta^2} \sin^2\left(\frac{\omega t}{2}\right)$, 
and its time average is \wzy{(with $\omega T\gg 1$)}
\begin{align}
\langle P_{e}^{\text{AE}}\rangle & \equiv \frac{1}{T}\int_0^T P_{e}^{\text{AE}} dt = \frac{1}{2}\frac{\Omega^2}{\Delta^2}\left[1-\frac{\sin(\omega T)}{\omega T}\right] \nonumber \\
& \approx \frac{1}{2}\frac{\Omega^2}{\Delta^2}, \label{eq:PeAEave}
\end{align}
which is inversely proportional to the square of the detuning $\Delta$. 
However, as shown in Eq.~\eqref{eq:HEff}, the effective control amplitude $\frac{\omega-\Delta}{2}\approx\Omega^2/(4\Delta)$ becomes small when the detuning $\Delta$ is large. For example, when $\Omega_1=\Omega_2$, we have the effective Hamiltonian 
$H_{\text{eff}}\approx\frac{\Omega^2}{4\Delta}\frac{(\sigma_x + I)}{2}$, where the Pauli operator $\sigma_x=|0\rangle\langle 1|+|1\rangle\langle 0|$ and $I$ is the identity operator, for state transitions between the states  $|0\rangle$ and $|1\rangle$ with a Rabi frequency $\Omega^2/(4\Delta)$.

\wzy{Without leakage to $|e\rangle$, an ideal implementation of Hamiltonian Eq.~\eqref{eq:HEff} realizes the gate within the subspace of $\{|0\rangle,|1\rangle\}$ as
\begin{align}
e^{-i H_{\text{eff}} t} = e^{i(\Delta-\omega)t/2}|b\rangle\langle b|+|d\rangle\langle d|. \label{eq:UGateGeneralBD}
\end{align}
This can be used to implement non-trivial quantum gates. For example, when $\Omega_1$ and $\Omega_2$ have the same amplitude but a different relative phase $\xi$, we have $|b\rangle=\frac{1}{\sqrt{2}}(|0\rangle+e^{-i\xi}|1\rangle)$ and
\begin{align}
e^{-i H_{\text{eff}} t}= R_\xi(\vartheta) \equiv \exp\left(-i\vartheta \frac{\sigma_\xi}{2}\right), \label{eq:UGateGeneral}
\end{align}
up to a trivial global phase. Here $\vartheta=(\omega-\Delta)t/2$ and $\sigma_\xi = \cos\xi \sigma_x + \sin\xi \sigma_y$ with $\sigma_y=-i |0\rangle\langle 1| + i|1\rangle\langle 0|$.
}

\section{Reducing the leakage in adiabatic elimination}

From the geometric picture in Fig.~\ref{fig:fig1}(c), we find that adding simple phase shifts to the control fields 
can significantly suppress the population in the intermediate state $|e\rangle$ during the control. See Fig.~\ref{fig:fig1}(b) for the simple modification on the AE scheme. We assume the same amplitudes $\Omega_1$ and $\Omega_2$ of the control fields as those in the AE scheme for a fair comparison between AE and our scheme.  

During the time $t<t_c$, we set the common phase $\phi(t)=0$ and the detuning $\Delta(t)=\Delta$ in the control Hamiltonian, as in the AE scheme. See Fig.~\ref{fig:fig1}(b). When the time $t_c$ for this control satisfies 
\begin{align}
\cos \phi_c=\Delta/(\Delta+\omega),
\end{align}
with 
\begin{align}
\phi_c\equiv\omega t_c<\pi
\end{align} 
for the smallest $t_c$, the state given by Eq.~\eqref{eq:psiBE} at $t=t_c$  becomes
\begin{align}
|\psi_{b,e}(t_c)\rangle  =|\varphi_{+}(t_c)\rangle,
\end{align}
which, nicely, is the eigenstate
\begin{align}
|\varphi_{+}(t_c)\rangle  = \cos\left(\frac{\phi_c}{2}\right)|b\rangle- \frac{i \sin(\frac{\phi_c}{2})}{\omega}(\Delta |b\rangle +\Omega |e\rangle),
\end{align}
of $\tilde{H}$ with $\phi=\phi_c$. See Fig\wzy{s}.~\ref{fig:fig1}(c) \wzy{and (d)}, where
the state $|b\rangle$ at the point $A$ evolves to the state at the point $B$. 

Then for the time $t\geq t_c$, we change the common phase to $\phi=\phi_c$ in the control Hamiltonian and keep the detuning unchanged. Since $|\psi_{b,e}(t_c)\rangle$ is an eigenstate of the control Hamiltonian,
the state $|\psi_{b,e}(t)\rangle=e^{-i\frac{\omega}{2}(t-t_c)}|\varphi_{+}(t_c)\rangle$ at a later time $t>t_c$  remains stationary on the Bloch sphere. 

The state $|\psi(t)\rangle$ [see Eq.~\eqref{eq:psiT}] at $t\geq t_c$ for $\Delta\gg\Omega$ reads
\begin{align}
|\psi(t)\rangle \approx e^{-i H_{\text{eff}} t}|\psi(0)\rangle + \alpha e^{-i(\phi_c+\omega t)}\frac{\Omega}{2\Delta} |e\rangle, \label{eq:psiOur}
\end{align}
where the effective Hamiltonian $H_{\text{eff}}$ for the subspace spanned by $|0\rangle$ and $|1\rangle$ is the same as the one [Eq.~\eqref{eq:HEff}] of conventional AE scheme. Notably, the population of the intermediate state $|e\rangle$ becomes
$P_{e}=\frac{1}{4}\frac{\Omega^2}{\Delta^2}$
and its time average
\begin{align}
\langle P_{e}\rangle=\frac{1}{4}\frac{\Omega^2}{\Delta^2}, \label{eq:PeOurAve}
\end{align}
is only half of the leakage in the conventional AE scheme given by Eq.~\eqref{eq:PeAEave}. In this manner,
adding a simple phase shift to the control field can greatly reduce the population in the intermediate \wzy{state.}

It is possible to further reduce the population of the intermediate state by adding another phase shift on the control fields as shown in Fig.~\ref{fig:fig1}(b). At the end of the control, we apply the control Hamiltonian for a 
time duration $t_c$ with $\phi=\pi$ and a change of the sign for the detuning, i.e., $\Delta \rightarrow -\Delta$. 
This will transfer $|\psi_{b,e}(t)\rangle \propto |\varphi_{+}(t_c)\rangle$, which has population in the $|e\rangle$ state, back to $|b\rangle$. In this manner, we have the exact result
\begin{align}
|\psi(t)\rangle =e^{-i (t-2t_c) H_{\text{eff}} }|\psi(0)\rangle.
\end{align}
with vanishing population in the $|e\rangle$ state and the Hamiltonian given by Eq.~\eqref{eq:HEff}. 
Note that in our case, the Hamiltonian for the control on the subspace spanned by $|0\rangle$ and $|1\rangle$ 
is exactly given by $H_{\text{eff}}$, Eq.~\eqref{eq:HEff}. Therefore, we do not require the condition $\Omega \ll \Delta$, which allows to greatly enhance the control amplitude and accelerate the control speed.

\begin{figure}[]
	\centering
	\includegraphics[width=\columnwidth]{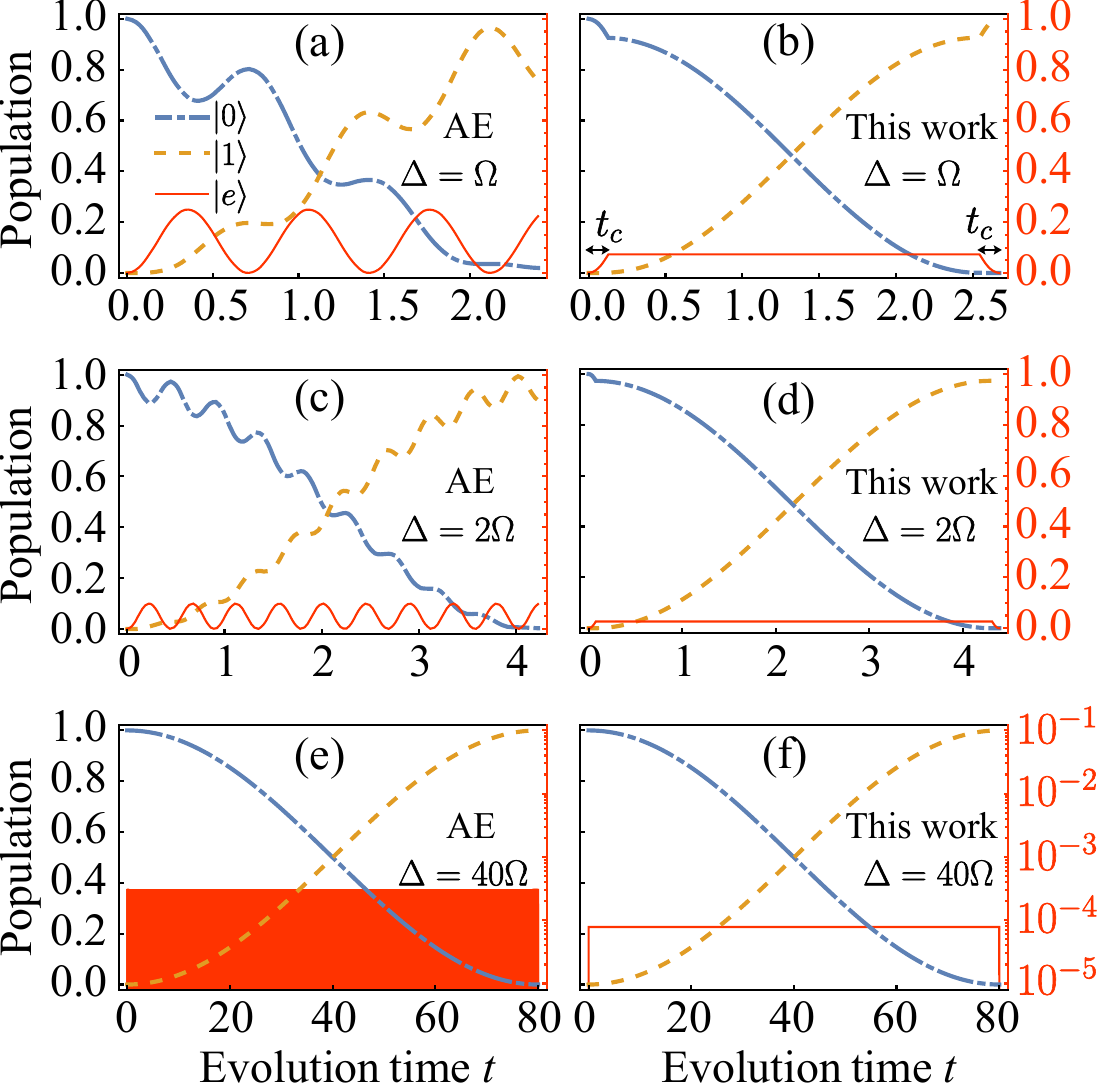}
	\caption{	Population transfer for AE (a\why{, c, e}), \wzy{and our scheme} (\why{b, d, f}) for various $\Delta$. \wzy{The blue dot-dashed lines (yellow dashed lines) represent the populations of $|0\rangle$ ($|1\rangle$), with their corresponding axis shown on the left side of the figures. The red solid thin line indicates the population of $|e\rangle$, with its axis displayed in red on the right side.} The parameters for the plots are $\Omega_1=\Omega_2$ and $\Omega=2\pi$. The evolution time is $2\pi/(\omega-\Delta)$ and $2\pi/(\omega-\Delta)+2t_c$ for AE and our scheme, respectively.}
	\label{fig:figAE}
\end{figure}

\begin{figure}[]
	\centering
	\includegraphics[width=\columnwidth]{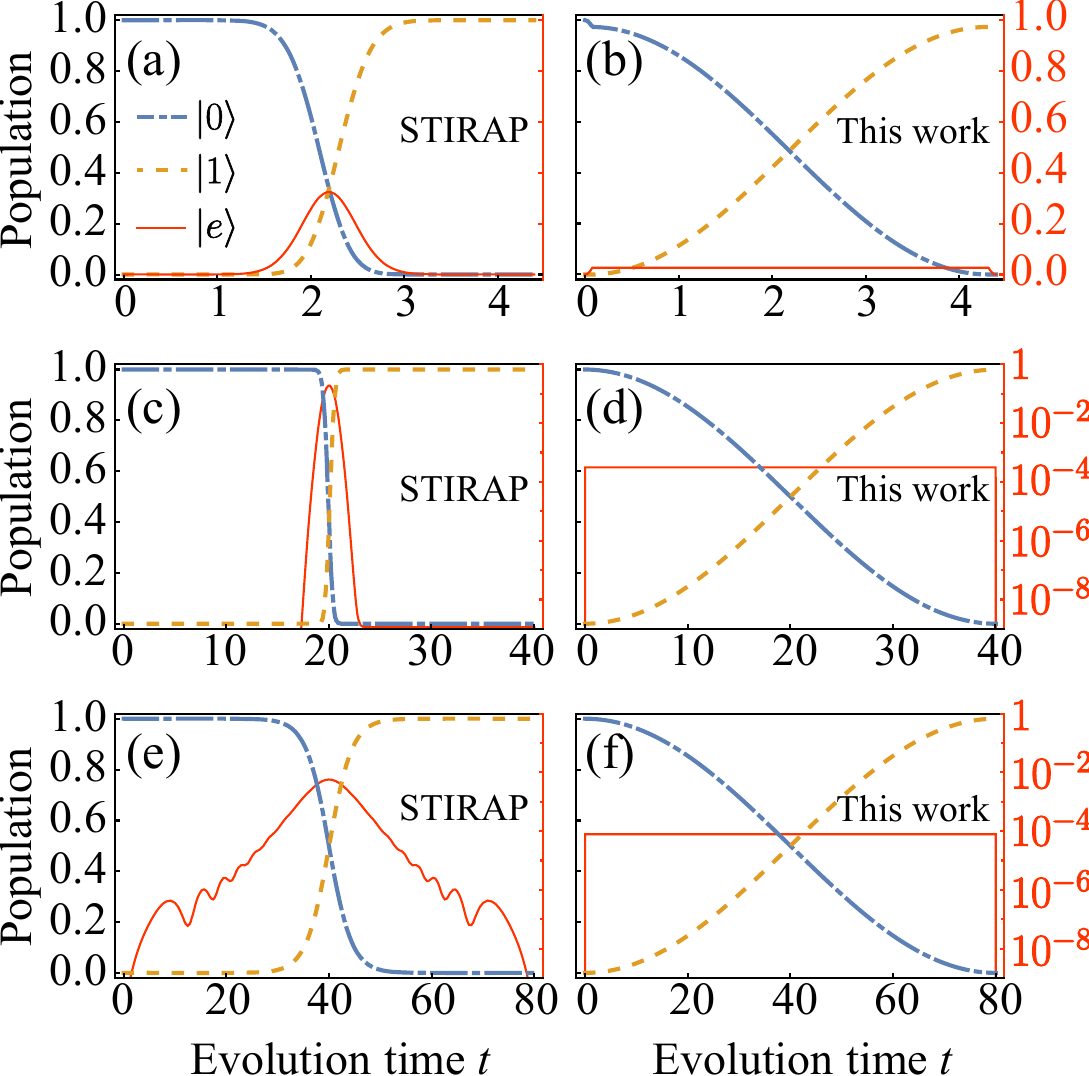}
	\caption{\why{Population transfer for STIRAP (a, c, e), and our scheme (b, d, f) is shown for various total evolution times. The blue dot-dashed lines (yellow dashed lines) represent the populations of $|0\rangle$ ($|1\rangle$), with their corresponding axes shown on the left side of the figures. The red solid thin line indicates the population of $|e\rangle$, with its axis displayed in red on the right side. The parameters of STIRAP in (a), (c), and (e) are $\{\sigma, t_m\}\approx \{0.75, 0.175\}$, $\{1.02, 0.4\}$, and $\{15.6, 13.4\}$, respectively. For our scheme in (b), (d), and (f), we set $\Delta=2\Omega$, $20\Omega$, and $40\Omega$, with $\Omega=2\pi$, such that the total evolution times match those used for STIRAP.}}
	\label{fig:figSTIRAP}
\end{figure}

In Fig.~\ref{fig:figAE}, we compare the performance of our scheme with the conventional AE scheme. The populations of state $|0\rangle$, $|1\rangle$\wzy{,} and $|e\rangle$ as a function of evolution time for AE and our method are shown in Fig.~\ref{fig:figAE}. It can be seen that our method not only reduces the population in the intermediate state $|e\rangle$ during the whole evolution process, but also perfectly evolves into the target state at the end. For the large detuning $\Delta\gg \Omega$, the population in the intermediate state $|e\rangle$\wzy{, approximately } $(\Omega/\Delta)^2 /2$\wzy{,} for AE method is adiabatically eliminated, but the effective coupling strength $\sim\Omega^2/\Delta$ between the two states $|0\rangle$ and $|1\rangle$ also becomes weaker.
\why{ We numerically calculate the average leakages in Fig.~\ref{fig:figAE}. The average leakage for AE is $1.20\times10^{-1}$ [Fig.~\ref{fig:figAE}(a)], $4.95\times10^{-2}$ [Fig.~\ref{fig:figAE}(c)], and $1.56\times10^{-4}$ [Fig.~\ref{fig:figAE}(e)], respectively, while the average leakage for our scheme are $6.86\times10^{-2}$ [Fig.~\ref{fig:figAE}(b)], $2.58\times10^{-2}$ [Fig.~\ref{fig:figAE}(d)], and $7.81\times10^{-5}$ [Fig.~\ref{fig:figAE}(f)], respectively.}
The maximum population of the intermediate state $|e\rangle$ in our method is only $1/4$ of that of AE method, and the average population in the intermediate state $|e\rangle$ in our method is only about half of that of AE method.

\wzy{Our scheme also exhibits a lower average leakage $\overline{P}_e$ compared to STIRAP, as shown in Fig.~\ref{fig:figSTIRAP}. The Hamiltonian of STIRAP takes the same form as Eq.~\eqref{eq:systemHamiltonian}, with the replacement  $\Omega_1 \rightarrow \Omega_1 \exp{\left[-(t-T/2-t_m)^2/\why{\sigma}^2\right]}$ and $\Omega_2 \rightarrow \Omega_2 \exp{\left[-(t-T/2+t_m)^2/\why{\sigma}^2\right]}$
for the two resonant adiabatic pulses. Here $T$ is the total time for the control.}
We keep the maximum coupling strength and evolution time the same as our method parameters, and search for the optimal parameters $t_m$ and \why{$\sigma$} that make the final state of STIRAP with the highest fidelity. As shown in \why{Fig.~\ref{fig:figSTIRAP}, the average leakage of our method is smaller that that of STIRAP.
We numerically calculate the average leakages in Fig.~\ref{fig:figSTIRAP}. The average leakage for STIRAP is $5.39\times10^{-2}$ [Fig.~\ref{fig:figSTIRAP}(a)], $4.89\times10^{-3}$ [Fig.~\ref{fig:figSTIRAP}(c)], and $5.32\times10^{-4}$ [Fig.~\ref{fig:figSTIRAP}(e)], respectively, while the average leakage for our scheme are $2.58\times10^{-2}$ [Fig.~\ref{fig:figSTIRAP}(b)], $3.12\times10^{-4}$ [Fig.~\ref{fig:figSTIRAP}(d)], and $7.81\times10^{-5}$ [Fig.~\ref{fig:figSTIRAP}(f)], respectively.}

\begin{figure}[]
\centering
\includegraphics[width=\columnwidth]{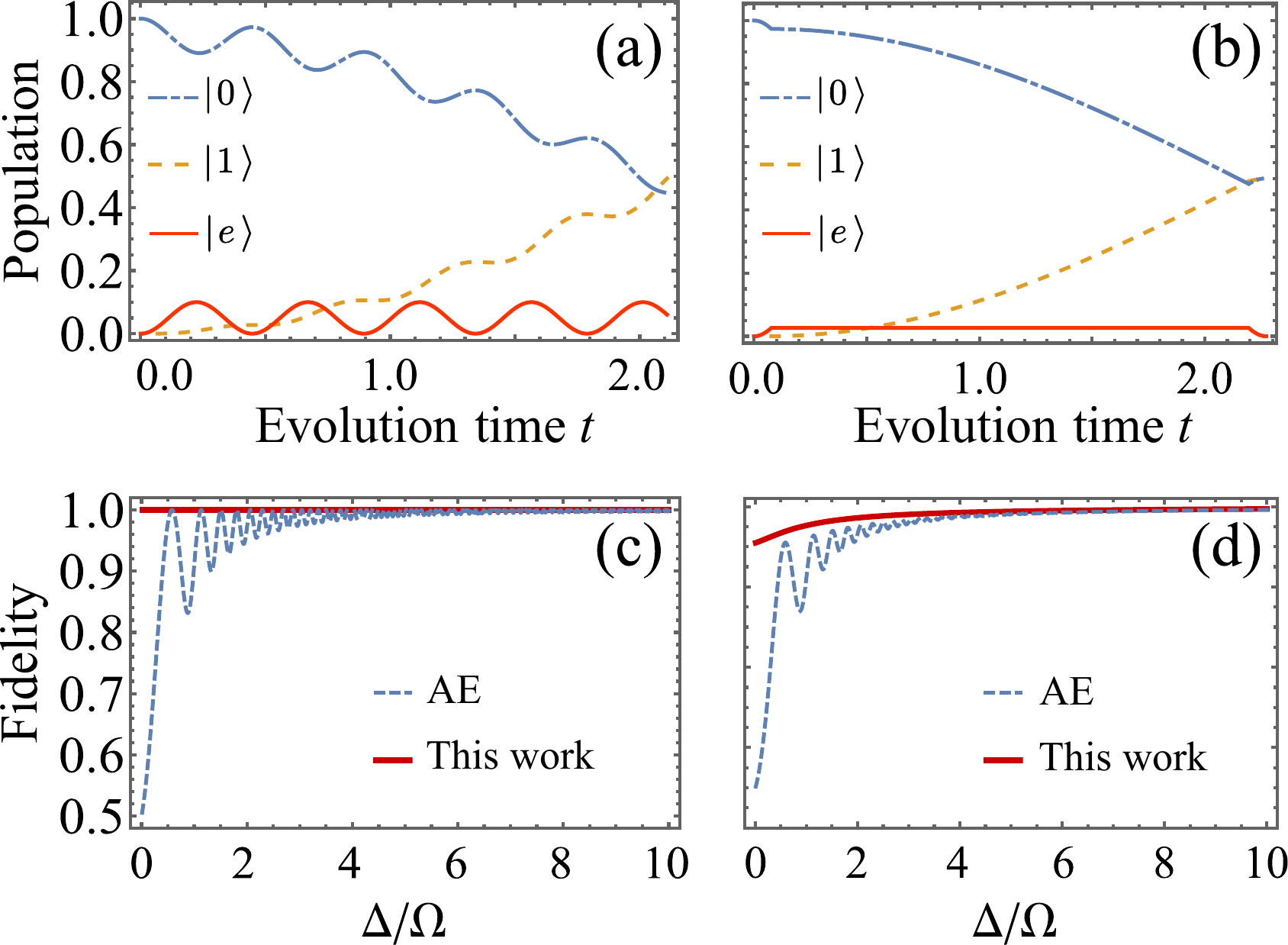}
\caption{The populations as a function of evolution time under the $R_{\text{x}}(\frac{\pi}{2})$ gate for AE (a) and our scheme (b). (c) The gate fidelity of $R_{\text{x}}(\frac{\pi}{2})$ as a function of $\Delta/\Omega$ without decoherence. (d) The state fidelity of initial state $|0\rangle$ under the $R_{\text{x}}(\frac{\pi}{2})$ as a function of $\Delta/\Omega$ with a decay rate $\gamma=0.5$. Here $\Omega_1=\Omega_2$, $\Omega=2\pi$, \wzy{and} $\Delta=2\Omega$. The evolution time is $\pi/(\omega-\Delta)$ for AE and $\pi/(\omega-\Delta)+2t_c$ for our scheme, respectively.}
\label{fig:figDecay}
\end{figure}

\wzy{Because of a lower leakage, our} scheme can be used to construct quantum gates \wzy{with higher fidelities}. In contrast to AE method, our method enables the initial state to evolve into the target subspace without leakage into the intermediate state $|e\rangle$, and the average leakage into the intermediate state $|e\rangle$ is reduced. Without loss of generality, we consider single-qubit gate $\wzy{R_{\xi=0}(\vartheta)\equiv} R_{\text{x}}(\vartheta)=\exp[-i\frac{\vartheta}{2}\sigma_x]$ \wzy{[see Eq.~\eqref{eq:UGateGeneral}], since this is equivalent to other gates of different axes by changing the relative phase $\xi$ of $\Omega_1$ and $\Omega_2$.}
Figures~\ref{fig:figDecay} (a) and (b) show the state evolution of the $R_{\text{x}}(\frac{\pi}{2})$ gate for AE and our method, respectively. It can be seen that our method can prepare the initial state to an ideal superposition state with lower average leakage $\overline{P}_e$. 

Our scheme can evolve to the ideal target state for any parameter region $\Delta/\Omega$. Figure~\ref{fig:figDecay}(c) shows the fidelity of $R_{\text{x}}(\frac{\pi}{2})$ gate\wzy{, which is equivalent to a Hadamard gate,} for AE and our scheme with the parameter $\Delta/\Omega$. The gate fidelity is defined as $F_{g}=\frac{1}{2}|{\text{Tr}}(U^{\dagger}U_{\text{id}})|$, where $U$ is the evolution operator of the Hamiltonian $H(t)$ achieved by AE or our scheme, and $U_{\text{id}}$ is the ideal gate \wzy{in the subspace of $\{|0\rangle,|1\rangle\}$}, e.g.,  $U_{\text{id}}=R_{\text{x}}(\frac{\pi}{2})$ for the $R_{\text{x}}(\frac{\pi}{2})$ gate. One can see that the gate fidelity of our method is always 1 without considering decoherence \wzy{and control errors}, while the gate fidelity of AE method depends on the specific parameter $\Delta/\Omega$.

Moreover, our method outperforms AE method even in the presence of decay. In the simulation shown in Fig.~\ref{fig:figDecay}(d), we use the Lindblad master equation with the decay operators $\sqrt{\gamma}|0\rangle\langle e|$ and $\sqrt{\gamma}|1\rangle\langle e|$. The state fidelity is defined as $F=$Tr$[\rho_{\text{id}}\rho]$, where $\rho_{\text{id}}$ and $\rho$ are the ideal density operator without decay and the density operator in the presence of dissipation, respectively. As shown in Fig.~\ref{fig:figDecay}(d), the fidelity of our method is higher than that of the AE method across different parameter regions of $\Delta/\Omega$. 

\begin{figure}[]
	\centering
	\includegraphics[width=\columnwidth]{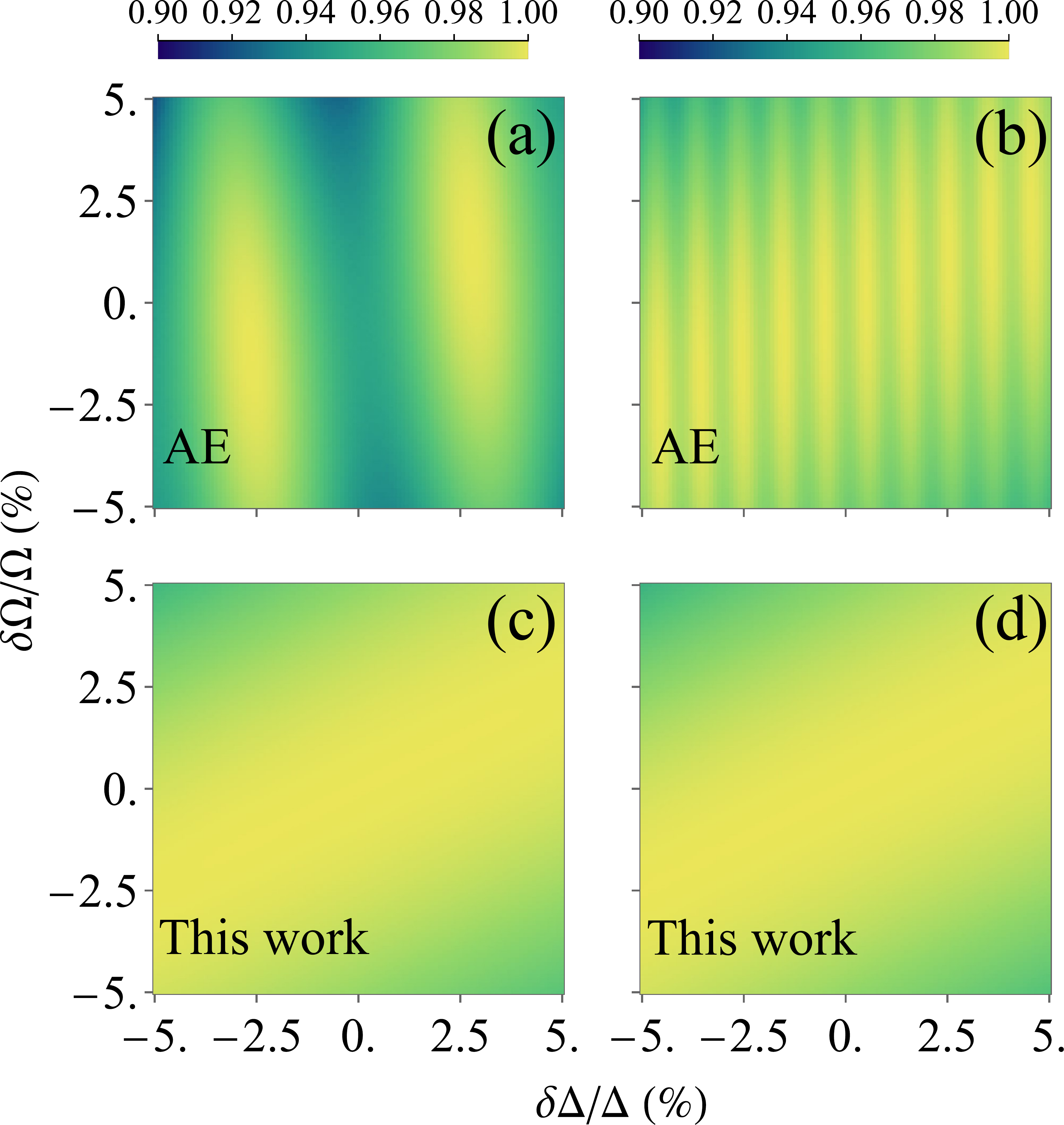}
	\caption{\wzy{The fidelities of $R_{\text{x}}(\pi)$ gate for AE [(a),(b)] and our scheme [(c),(d)] in the presence of amplitude error $\delta\Omega$ and detuning error $\delta\Delta$. (a) and (c) $\Delta/\Omega=3$. (b) and (d) $\Delta/\Omega=7$. Here $\Omega_1=\Omega_2$ and $\Omega=2\pi$.}}
	\label{fig:figRob}
\end{figure}

\begin{figure}[]
	\centering
	\includegraphics[width=\columnwidth]{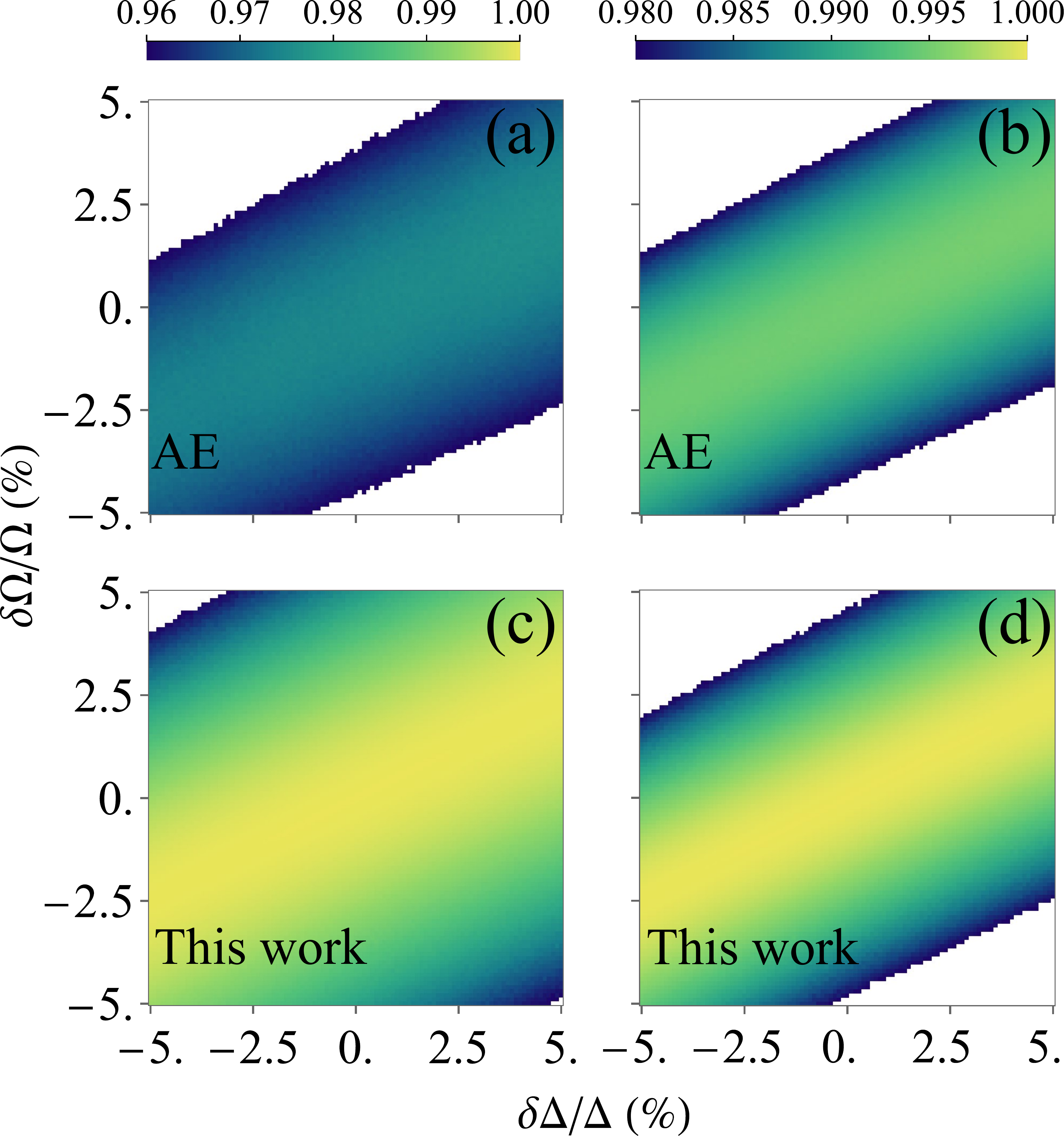}
	\caption{\why{The same as Fig.~\ref{fig:figRob}, but for the average fidelities of $R_{\text{x}}(\vartheta)$, averaged over $\vartheta\in(0,2\pi]$.}}
	\label{fig:figRobAve}
\end{figure}

\wzy{We consider the static amplitude error $\delta\Omega$ and the static detuning error $\delta\Delta$, which modify the coupling strength and the detuning to $\Omega+\delta\Omega$ and $\Delta+\delta\Delta$, respectively. In Appendix~\ref{sec:controErrors} we derive the leading-order approximated leakage to the intermediate state $|e\rangle$ for the conventional AE method,
\begin{align}
P_{e}^{\text AE}(t)\approx|\alpha|^2 \frac{\Omega ^2}{\Delta ^2}\sin^2\left(\frac{\omega t}{2}+\frac{t \delta\Omega  \sin\theta}{2}\right),
\label{eq:PeAE-Amp}
\end{align}
when there is a static amplitude error $\delta\Omega$, and  
\begin{align}
P_{e}^{\text AE}(t) \approx|\alpha|^2 \frac{\Omega ^2}{\Delta ^2}\sin^2\left(\frac{\omega t}{2}+\frac{t \delta\Delta\cos\theta }{2}\right),
\label{eq:PeAE-detuning}
\end{align}
in the presence of detuning error $\delta\Delta$. Here $\alpha=\langle b|\psi(0)\rangle$ represents the $|b\rangle$ state component of the initial state $|\psi(0)\rangle$. For our scheme, the leading-order approximated leakage $P_{e}(t)\approx 0$ is insensitive to control errors $\delta\Omega$ and $\delta\Delta$, see Appendix~\ref{sec:controErrors}. }

\wzy{
We show in Fig.~\ref{fig:figRob} the fidelities of an $R_{\text{x}}(\vartheta)$ gate with $\vartheta=\pi$, using a uniform averaging over all possible initial states in the subspace spanned by $\{|0\rangle,|1\rangle\}$.
As shown in Fig.~\ref{fig:figRob}, with the increase of $\Delta/\Omega$, the fidelities of AE become better since 
the population in the intermediate state $|e\rangle$ is reduced. On the contrary, our scheme provides higher fidelities 
over a board range of parameters. The periodic pattern in Fig.~\ref{fig:figRob} for the AE method is consistent with Eqs.~\eqref{eq:PeAE-Amp} and \eqref{eq:PeAE-detuning} (note that $\sin\theta\approx \Omega/\Delta$ is smaller than $\cos\theta$). 
}

\wzy{
We also show in Fig. \ref{fig:figRobAve} the average fidelity of the $R_{\text{x}}(\vartheta)$ gate for all $\vartheta\in[0,2\pi)$ by using Monte Carlo sampling of $10^4$ randomly generated $\vartheta$. In Appendix~\ref{sec:controErrors} we derive the corresponding average leakage. For the conventional AE method, the average leakage reads
\begin{align}
\langle{P_{e}^{\text AE}}\rangle_\vartheta
&\approx \frac{1}{2}|\alpha|^2 \frac{\Omega ^2}{\Delta ^2}, \label{eq:PeAve}
\end{align}
which is consistent with Fig. \ref{fig:figRobAve}. As shown in Fig. \ref{fig:figRobAve}, the average gate fidelity for our method is higher than that obtained by AE method in the presence of errors. 
}

\wzy{\section{Example: application to the control for two-qubit gates\label{sec:qubit}}}
Here we illustrate an example of our method for reducing leakage errors in the implementation of two-qubit gates. We consider the Hamiltonian $H^\prime=H_0+H_1$, with 
$H_0=\omega a^{\dagger}a+\sum_{k=1}^{2}\frac{\omega_k}{2}\sigma^{z}_{k}$ and $H_1=\sum_{k=1}^{2} g_{k}\sigma^{+}_{k}a e^{i\phi(t)} + \text{H.c.}$~\cite{fink2009}.
Here $a^{\dagger} (a)$ is the creation (annihilation) operator for a cavity field with a frequency of $\omega$, $\omega_k$ is the energy difference between state $|\uparrow\rangle$ and $|\downarrow\rangle$ for qubit $k$, $\sigma^{+}_{k}(\sigma^{-}_{k})$ is the raising (lowering) operator for qubit $k$, and $g_{k}$ is the coupling strength between the qubit $k$ and the cavity.

In the interaction picture with respect to $H_0$, the Hamiltonian becomes
\begin{align}
H_{I}=\sum_{k=1}^{2}\left[\Delta(t)\frac{\sigma^{z}_{k}}{2}+g_{k}\sigma^{+}_{k}a e^{i\phi(t)} + \text{H.c.}\right],
\end{align}
where we have set the detuning $\Delta_{k}=\omega_{k}-\omega =\Delta(t)$. We denote $|0_c\rangle$ and $|1_c\rangle$ as the ground and first excited states of the cavity field ($\omega a^\dagger a$). The subspace spanned by $\{|0_c\downarrow \uparrow\rangle$, $|0_c\uparrow \downarrow\rangle$, $|1_c\downarrow\downarrow\rangle\}$, which contains only one excitation, is decoupled from other states, and we have the Hamiltonian within this subspace~\cite{zhang2022},
\begin{align}
H=-\Delta(t)|e\langle e| + \left(\frac{\Omega}{2}e^{i\phi(t)}|b\rangle\langle e| + \text{H.c.}\right),
\end{align}
which has the same form as Eq.~\eqref{eq:systemHamiltonian} with $|e\rangle \equiv |1_c\downarrow\downarrow\rangle$, $|b\rangle \equiv \frac{2}{\Omega}(g_1|0_c\uparrow \downarrow\rangle+g_2|0_c\downarrow \uparrow\rangle)$, and $\Omega=2\sqrt{|g_1|^2+|g_2|^2}$.

Note that the leakage to the state $|e\rangle=|1_c\downarrow\downarrow\rangle$ induces decoherence to the 
qubits because it has a different cavity excitation number than the other two basis states $\{|0_c\downarrow \uparrow\rangle$, $|0_c\uparrow \downarrow\rangle\}$. Using our scheme, we can efficiently suppress the leakage to the state $|e\rangle = |1_c\downarrow\downarrow\rangle$, and thereby, resulting in a high-fidelity control for the qubits. The resulting control 
\begin{align}
H_{\text{eff}} = \frac{\omega-\Delta}{2}|b^\prime\rangle\langle b^\prime|, 
\end{align}
with $|b^\prime\rangle =  \frac{2}{\Omega}(g_1|\uparrow \downarrow\rangle + g_2|\downarrow \uparrow\rangle)$, can be used to construct high-fidelity two-qubit gates on the two qubits with the state space $\{ |\uparrow \uparrow\rangle, |\uparrow \downarrow\rangle,|\downarrow \uparrow\rangle,|\downarrow \downarrow\rangle\}$.

~

\section{Conclusion}
In conclusion, we have proposed a simple quantum control scheme by means of phase shifts and changing the detuning in two-photon resonance three-level systems. Our scheme can achieve the ideal quantum gate operations between two target state, with the average population leakage to the intermediate state reduced to approximately half that of the conventional AE method.
Furthermore, even in the presence of decoherence, the fidelity of our scheme across various parameter regimes of $\Delta/\Omega$ is higher than that of the AE method. 
When considering static amplitude and detuning errors, our scheme is more robust than the AE method. Our scheme can be applied to various quantum systems, such as ion traps~\cite{Leibfried2003IonRMP}, neutral atoms~\cite{Saffman2010RMP}, superconducting qubits~\cite{Blais2021CQEDRMP}, and solid-state defects such as nitrogen-vacancy color centers~\cite{Doherty2013Review}. We anticipate that the ideas underlying our scheme can be extended to more general settings, such as the construction of high-fidelity multi-qubit gates for quantum information processing~\cite{blais2020QIP} and high-fidelity time-dependent control for quantum sensing~\cite{Degen2017RMPsensing,xu2024enhancing,zeng2024wide}.

\begin{acknowledgments}
This work was supported by National Natural Science Foundation of China (Grant No.~12074131) and the Natural Science Foundation of Guangdong Province (Grant No.~2021A1515012030\wzy{)}. H.-Y.W. and X.-Y.C. contributed equally to this work.
\end{acknowledgments}

\appendix

\wzy{\section{Perturbative analysis on the effect of control errors \label{sec:controErrors}}}
In this appendix, we analyze the performance of the control methods when there are static control errors.

A static amplitude fluctuation, i.e., $\Omega \rightarrow \Omega+\delta\Omega$ introduces to the original AE Hamiltonian [see Eq.~\eqref{eq:systemHamiltonian}] an amplitude error term
\begin{align}
	H_{\delta\Omega}=\frac{\delta\Omega}{2}\left(e^{i\phi}|b\rangle\langle e| + \text{H.c.}\right).
\end{align}
And therefore, the total Hamiltonian becomes
\begin{align}
H_{A}=H_0+H_{\delta\Omega},
\label{HdeltaOmega}
\end{align}
where, written in the diagonal form, 
\begin{align}
H_0 &=0|d\rangle\langle d|+ E_+|+\rangle\langle+|+E_-|-\rangle\langle-|,
\end{align}
with the eigenenergies $E_+ =(\omega-\Delta)/2$ and $E_-=-(\omega+\Delta)/2$, and the eigenstates
\begin{align}
|+\rangle=-i\cos\left(\frac{\theta}{2}\right)e^{i\phi}|b\rangle-i\sin\left(\frac{\theta}{2}\right)|e\rangle,
\end{align}
and 
\begin{align}
|-\rangle=-i\sin\left(\frac{\theta}{2}\right)|b\rangle+i\cos\left(\frac{\theta}{2}\right)e^{-i\phi}|e\rangle.
\end{align}
We treat $H_{\delta\Omega}$ as a perturbation ($\delta\Omega \ll |E_+|, |E_-|$) and keep only the leading-order terms using the time-independent perturbation theory. That is, under the secular approximation, we write
\begin{align}
H_{\delta\Omega} \approx \frac{\delta\Omega\sin\theta}{2}\left(|+\rangle\langle+|-|-\rangle\langle-|\right),
\label{HIdeltaOmega}
\end{align}
which commutes with $H_0$. Therefore, we obtain
\begin{align}
H_{A} \approx &   0|d\rangle\langle d|+(-\frac{\Delta}{2}+\frac{\omega}{2}+\frac{\delta\Omega}{2}\sin\theta)|+\rangle\langle+| \notag\\
& -(\frac{\Delta}{2}+\frac{\omega}{2}+\frac{\delta\Omega}{2}\sin\theta)|-\rangle\langle-|,
\label{ap:eq:HAapprox}
\end{align}
and the corresponding evolution operator
\begin{align}
U_A(t) = e^{-i H_A t}.\label{ap:eq:UA}
\end{align}

Using Eqs.~\eqref{ap:eq:HAapprox} and \eqref{ap:eq:UA}, we calculate the leakage to the intermediate state $|e\rangle$ for any initial state $|\psi(0)\rangle=\alpha|b\rangle+\beta|d\rangle$. 

For conventional AE method, the control phase $\phi$ (e.g., $\phi=0$) is a constant, we obtain the leakage to the intermediate state $|e\rangle$ in the main text [Eq.~\eqref{eq:PeAE-Amp}]
\begin{align}
P_{e}^{\text AE}(t)&\equiv|\langle e|U_A(t)|\psi(0)\rangle|^2\notag\\
&\approx4|\alpha|^2\sin^2\frac{\theta}{2}\cos^2\frac{\theta}{2}\sin^2\left(\frac{\omega t}{2}+\frac{t \delta\Omega  \sin\theta}{2}\right)\notag\\
&=|\alpha|^2 \sin^2\theta\sin^2\left(\frac{\omega t}{2}+\frac{t \delta\Omega  \sin\theta}{2}\right)\notag\\
&\approx|\alpha|^2 \frac{\Omega ^2}{\Delta ^2}\sin^2\left(\frac{\omega t}{2}+\frac{t \delta\Omega  \sin\theta}{2}\right),
\end{align}
where $\alpha=\langle b|\psi(0)\rangle$. 

We consider the average performance of the gate $R_{\text x}(\vartheta)$ for all possible $\vartheta\in(0,2\pi]$, with $\vartheta =(\omega-\Delta)t/2$. And because $\omega-\Delta\ll \omega$, $\omega t$ varies from  $\omega t =0$ to  $\omega t \gg 1$ when $\vartheta$ varies from $\vartheta=0$ to $\vartheta=2\pi$. This gives the average of $P_{e}^{\text AE}(t)$ for different $\vartheta\in(0,2\pi]$:
\begin{align}
\langle{P_{e}^{\text AE}}\rangle_\vartheta
&\approx \frac{1}{2}|\alpha|^2 \sin^2\theta \approx \frac{1}{2}|\alpha|^2 \frac{\Omega ^2}{\Delta ^2}, \label{ap:eq:PeAve}
\end{align}
which is Eq.~\eqref{eq:PeAve} in the main text.

For our scheme, we use a short driving time $t_c$ to transfer the initial state $|\psi(0)\rangle$ to an eigenstate of the subsequent Hamiltonian with a phase shift of $\phi_c$. The effect of control error during the first part of the control (with $t\leq t_c$) can be neglected since $\delta\Omega t_c \ll 1$. With a phase shift $\phi_c$ in $H_A$ and setting the initial state to be an eigenstate of $H_A$, we obtain the leakage to the intermediate state $|e\rangle$ during the second part of our control
\begin{align}
P_{e}(t) \approx |\alpha|^2\sin^2\frac{\theta}{2}\approx \frac{1}{4}|\alpha|^2 \frac{\Omega ^2}{\Delta ^2},
\label{eq:PeOurA}
\end{align}
which is independent of $\delta\Omega$ in the leading-order approximation. After the detuning change and the second phase shift of our scheme, we eliminate the leakage to the  intermediate state $|e\rangle$ at the end of the control, i.e.,
\begin{align}
P_{e}(t)\approx 0,
\end{align}
in the leading-order approximation.

We perform the same perturbative analysis for the case of detuning error, which introduces the detuning error term
\begin{align}
H_{\delta\Delta}=-\delta\Delta|e\rangle\langle e|.
\end{align}
Under the secular approximation, 
\begin{align}
H_{\delta\Delta}\approx -\delta\Delta\left(\sin^2\frac{\theta}{2}|+\rangle\langle+|+\cos^2\frac{\theta}{2}|-\rangle\langle-|\right).
\end{align}
For the conventional AE method, the leakage in the presence of detuning error is
\begin{align}
P_{e}^{\text AE}(t) & \approx 4|\alpha|^2\sin^2\frac{\theta}{2}\cos^2\frac{\theta}{2}\sin^2\left(\frac{\omega t}{2}+\frac{t \delta\Delta\cos\theta }{2}\right)\notag\\
& =|\alpha|^2 \sin^2\theta\sin^2\left(\frac{\omega t}{2}+\frac{t \delta\Delta\cos\theta }{2}\right) \notag\\
& \approx|\alpha|^2 \frac{\Omega ^2}{\Delta ^2}\sin^2\left(\frac{\omega t}{2}+\frac{t \delta\Delta\cos\theta }{2}\right).
\end{align}
For this case, the average of $P_{e}^{\text AE}(t)$ for different $\vartheta\in(0,2\pi]$ is also given by Eq.~\eqref{ap:eq:PeAve}. 

For our scheme, we find that the leakage in the presence of detuning error takes the same form as the leakage in the presence of amplitude error, using the leading-oder approximation as we treated in the conventional AE method. 

~

%

\end{document}